\documentclass[aps,10pt,prb,twocolumn,showpacs,floatfix]{revtex4-1}

\usepackage{verbatim}
\usepackage{graphicx}
\usepackage{dcolumn}
\usepackage{bm}
\usepackage{color}
\usepackage{hyperref}
\usepackage{makeidx}
\usepackage{amsmath}
\makeindex
\def\>{\right\rangle}
\def\<{\left\langle}
\def\be{\begin{equation}}
\def\ee{\end{equation}}
\def\ba{\begin{array}{l}}
\def\ea{\end{array}}

\def\beq{\begin{eqnarray}}
\def\eeq{\end{eqnarray}}
\begin{document}
\title{Emission of entangled Kramers pairs from a helical mesoscopic capacitor}
\author{Giacomo Dolcetto and Thomas L. Schmidt}
\affiliation{Physics and Materials Science Research Unit, University of Luxembourg, L-1511 Luxembourg.
}
\date{\today}
\begin{abstract}
The realization of single-electron sources in integer quantum Hall systems has paved the way for exploring electronic quantum optics experiments in solid-state devices. In this work, we characterize a single Kramers pair emitter realized by a driven antidot embedded in a two-dimensional topological insulator, where spin-momentum locked edge states can be exploited for generating entanglement.
Contrary to previous proposals, the antidot is coupled to both edges of a quantum spin Hall bar, thus enabling this mesoscopic capacitor to emit an entangled two-electron state.
We study the concurrence $\mathcal{C}$ of the emitted state and the efficiency $\mathcal{F}$ of its emission as a function of the different spin-preserving and spin-flipping tunnel couplings of the antidot with the edges.
We show that the efficiency remains very high ($\mathcal{F}\geq 50\%$) even for maximally entangled states ($\mathcal{C}=1$). We also discuss how the entanglement can be probed by means of noise measurements and violation of the Clauser-Horne-Shimony-Holt inequality.
\end{abstract}

\pacs{73.23.-b, 03.65.Bg, 72.10.-d, 85.35.Gv}
\maketitle
\section{Introduction}\label{sec:intro}
Electron quantum optics can be regarded as the fermionic counterpart of standard quantum optics based on photons~\cite{Boquillon2014}.
The latter is built on three crucial ingredients: phase-coherent photon waveguides, beam splitters, and single-photons sources.
Therefore, to perform quantum-optics experiments with electrons, a huge effort has been invested to transfer these ingredients to solid-state devices.
In this respect, the edge states of quantum Hall systems provide suitable phase-coherent wave-guides for electrons, because transport is ballistic due to their intrinsic chirality~\cite{Grenier2011}.
Moreover, the electron counterpart of photon beam splitters is then naturally found in quantum point contacts (QPCs), which make it possible to mix and recombine the incoming electron fluxes~\cite{Ji2003, Neder2007}, just as the photon beam splitters separate the photon beam into transmitted and reflected components.
Finally, the single-electron source has been recently experimentally realized.
This has been achieved by means of driven mesoscopic capacitors (MCs)~\cite{Feve2007, Mahe2010} or Lorentzian voltage pulses~\cite{Dubois2013a, Dubois2013b}, thus accomplishing the receipt needed to implement electron quantum optics.

These achievements have paved the way for realizing fascinating experiments: quantum tomography protocols to measure single-electron decoherence,\cite{Jullien2014} the investigation of indistinguishability and fermionic statistics via antibunching effects in two-particle interferometric setups~\cite{Boquillon2012} and the detection of charge fractionalization in the presence of interactions\cite{Boquillon2013} represent notable examples.

Recently, two-dimensional topological insulators (2D TIs)~\cite{Hasan2010, Qi2011, Konig2007, Knez2011} have also been considered as an interesting playground for implementing electron quantum optics experiments~\cite{Ferraro2014electronic, Jonckheere2012, Hofer2013emission, Inhofer2013proposal, Xing2014, Chen2012, Strom2015controllable, Calzona2016}.
Here, two electron waveguides emerge on the edge, one for spin-up and one for spin-down electrons. However, contrary to standard one-dimensional systems, the bulk topological properties force the two species to propagate in opposite directions and time-reversal symmetry (TRS) prevents backscattering between the two channels~\cite{Wu2006}. Therefore, 2D TIs support phase-coherent ballistic transport on the edges~\cite{Dolcetto2016}. The role of beam splitters can be played by QPCs, and it is noteworthy that the range of possible applications is even richer than for QH systems because the incoming electrons have a larger number of possible scattering channels due to the additional spin degree of freedom~\cite{Sternativo2014}.
Furthermore, the mesoscopic capacitors implemented in 2D TIs inject pairs of electrons instead of single electrons because of Kramers degeneracy~\cite{Hofer2013emission}.
This richness leads to a novel antibunching phenomenon termed $\mathbf{Z}_2$ dip~\cite{Ferraro2014electronic, Edge2013} in contrast to the Pauli dip observed in QH channels~\cite{Wahl2014}, and can be exploited to create ac current sources which can be tuned to induce either pure charge or pure spin currents\cite{Inhofer2013proposal, Hofer2014}.
Moreover, the injection of two-particle (electrons or holes) states is very attractive for the creation and manipulation of entanglement in solid-state devices~\cite{Burkard2000, Recher2001, Ch2002, Bena2002, Beenakker2003, Leyati2007, Sato2010, Das2012, Schroer2014, Dasenbrook2015}, which is at the basis of quantum information processing~\cite{Nielsen2000quantum, Loss1998}, so that creation of entanglement in 2D TIs has been recently proposed~\cite{Hofer2013emission, Inhofer2013proposal, Sato2014, Chen2014, Strom2015controllable}.

\begin{figure}[t]
\centering
\includegraphics[width=8cm,keepaspectratio]{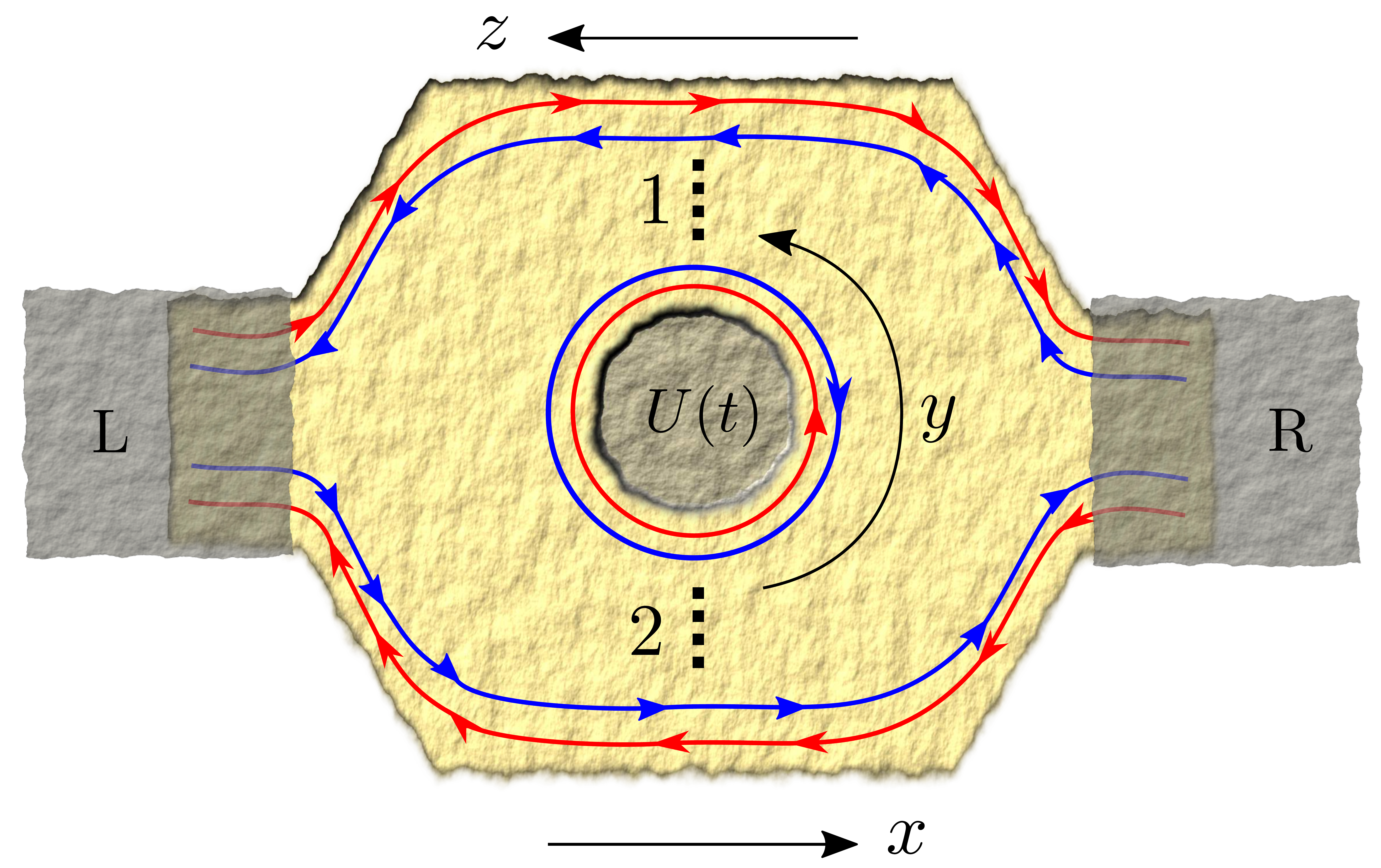}
\caption{(Color online) Antidot (central grey area) realized in a narrow QSH bar (yellow) coupled to both the two edges in a two-terminal configurations (left (L) and right (R) grey rectangles). Tunnel junction 2 is created at $x=0$, $y=0$ with tunnel junction 1 at $y=\pi R$, $z=0$. Spin up and spin down edge states are depicted in red and blue respectively. The antidot is driven by a time-dependent potential $U(t)$.}\label{fig:antidotQSHdouble}
\end{figure}

In this work we consider the device schematically shown in Fig.~\ref{fig:antidotQSHdouble}, which consists of a quantum spin Hall (QSH) antidot~\cite{Dolcetto2013, Hwang2014, Posske2013, Posske2014, Rod2016} acting as a MC~\cite{Buttiker1993, Moskalets2008}. The antidot can be realized either by mechanically etching the sample or, in the case of InAs/GaSb quantum wells, by gating the central region and thus causing a transition from QSH to trivial insulator~\cite{Liu2008}. In both cases, a pair of helical edge states appears around the antidot, in addition to the ones appearing at the external edges of the bar.
The antidot is tunnel-coupled to both the two edges of the QSH bar.
Contrary to the edge states, whose energy spectrum can be assumed to be continuous, the finite size of the antidot makes its spectrum discrete with an energy spacing $\Delta=v/R$, $v$ being the Fermi velocity and $R$ the radius of the antidot.
By applying a time-periodic gate voltage $U(t)$, its energy levels can be shifted periodically above and below the Fermi level of the edges, thus allowing the antidot to operate as a MC, able to periodically inject Kramers pairs of electrons or holes.
Contrary to previous proposals~\cite{Hofer2013emission, Inhofer2013proposal, Strom2015controllable, Ferraro2014electronic}, which always considered coupling with a single edge, we characterize the ability of the driven antidot to inject particles into both edges: one electron is injected into one edge, and its time-reversal partner is injected either into the same edge, with opposite spin because of Pauli principle, or in the other one, with arbitrary spin orientation. Therefore, we expect the injection process to be much richer than the one occurring when the MC is coupled to a single edge, in which case the only possibility in the presence of TRS is to inject the two electrons forming the Kramers pair into the same edge with opposite spin.
This richer scenario is particularly interesting when turning the attention to entanglement production.
Indeed, the injected state is in general a superposition of many different orthogonal states, and we show that entanglement production is possible only when the MC is coupled to both edges.
This modification also leads to peculiar transport properties from which we can extract information about the entanglement created.
In particular, by considering the two-terminal setup of Fig.~\ref{fig:antidotQSHdouble}, we find that if the Kramers pair is injected into a single edge then exactly one particle is collected in each detector, while the possibility to split the Kramers pair in the two edges give rise to alternative scenarios in which both particles are collected at the same contact.
Therefore, we are able to relate the concurrence, which measures the entanglement production, to the zero-frequency noise produced in a simple two-terminal configuration, thus providing a direct connection between quantum effects and standard transport measurements.
Furthermore, we show that the efficiency of the device, \textit{i.e.}, the ratio between the number of emitted entangled states and the total one, is very high compared to previous proposals.

The paper is organized as follows.
In Sec.~\ref{sec:MC}, we solve the dynamical scattering problem for the driven helical antidot coupled to the edges of the 2D TI. In particular, we compute the current injected in each channel and demonstrate that exactly one electron and one hole Kramers pair are injected in each cycle of the drive.
We also compute the zero-frequency noise and discuss the conditions under which the different channels are correlated.
In Sec.~\ref{sec:entanglement}, we introduce the concurrence $\mathcal{C}$ and the efficiency $\mathcal{F}$, which measure the ability of the device to generate entangled states.
We show that even though there is no entanglement at perfect efficiency ($\mathcal{C}=0$ for $\mathcal{F}=1$), the efficiency in the case of maximally entangled emitted states is very high compared to previous proposals, namely we find $\mathcal{F}=50\%$ for $\mathcal{C}=1$.
Finally, we establish a direct connection between these quantities and the zero-frequency noise measured in the two terminal configuration, which one can easily measure in experiments.
We also suggest an alternative protocol to detect the entanglement via violation of a Clauser-Horne-Shimony-Holt inequality~\cite{CHSH1969}.
Section~\ref{sec:conclusions} is devoted to the conclusions.

\section{The antidot as a mesoscopic capacitor}\label{sec:MC}
To characterize the MC we need to solve the dynamical scattering problem~\cite{Moskalets2011non} associated with the tunneling processes between the edge states and the driven antidot.
We define the scattering states on the edges and around the antidot as
\begin{eqnarray}
\psi(t,x)&=&e^{-i\frac{E}{\hbar}t}\times\left \{ \begin{matrix}
\left (\begin{matrix}B_{2\uparrow}\left (t+\frac{x}{v}\right )e^{-ikx} \\ A_{2\downarrow}e^{ikx}\end{matrix}\right ) & x<0 \\ \left (\begin{matrix}A_{2\uparrow}e^{-ikx} \\ B_{2\downarrow}\left (t-\frac{x}{v}\right )e^{ikx}\end{matrix}\right ) & x>0
\end{matrix} \right .\nonumber\\
\psi(t,y)&=&e^{-i\frac{E}{\hbar}t}\times\left \{\begin{matrix} \left (\begin{matrix}
c_{\uparrow}\left (t-\frac{y}{v}\right )\Upsilon(t)e^{iky} \\ c_{\downarrow}\left (t+\frac{y}{v}\right )\Upsilon(t)e^{-iky}
\end{matrix}\right ) &  0<y<\pi R \nonumber\\
\left (\begin{matrix}
d_{\uparrow}\left (t-\frac{y}{v}\right )\Upsilon(t)e^{iky} \\ d_{\downarrow}\left (t+\frac{y}{v}\right )\Upsilon(t)e^{-iky}
\end{matrix}\right ) &  \pi R<y<2\pi R
\end{matrix}\right .\\
\psi(t,z)&=&e^{-i\frac{E}{\hbar}t}\times\left \{ \begin{matrix}
\left (\begin{matrix} B_{1\uparrow}\left (t+\frac{z}{v}\right )e^{-ikz} \\ A_{1\downarrow}e^{ikz}\end{matrix}\right ) & z<0 \\ \left (\begin{matrix} A_{1\uparrow}e^{-ikz} \\ B_{1\downarrow}\left (t-\frac{z}{v}\right )e^{ikz} \\\end{matrix}\right ) & z>0
.\end{matrix} \right .\label{eq:wavefunctions}
\end{eqnarray}
Here $B_{i\sigma}$, $c_{\sigma}$ and $d_{\sigma}$ are scattering amplitudes, $A_{i\sigma}$ are the incoming amplitudes and $\Upsilon(t)=\exp[-(ie/\hbar)\int_{-\infty}^t dt' ~ U(t')]$ accounts for the phase acquired by the electron when moving in the time-dependent potential.
Note that, within our choice of coordinates $x,y,z$ (see Fig.~\ref{fig:antidotQSHdouble}), left-moving spin up and right-moving spin down electrons propagate on the edges, while right-moving spin up and left-moving spin down propagate around the antidot.
The wave-functions in Eq.~\eqref{eq:wavefunctions} are connected to each other through the scattering matrices of the QPCs as
\begin{equation}\label{eq:QPC1}
\left (\begin{matrix}
\psi_{\uparrow}(t,x=0^-) \\ \psi_{\downarrow}(t,y=2\pi R^-) \\ \psi_{\uparrow}(t,y=0^+) \\ \psi_{\downarrow}(t,x=0^+)
\end{matrix}\right )=S_2 \left (\begin{matrix}
\psi_{\downarrow}(t,x=0^-) \\ \psi_{\uparrow}(t,y=2\pi R^-) \\ \psi_{\downarrow}(t,y=0^+) \\ \psi_{\uparrow}(t,x=0^+)
\end{matrix}\right )
\end{equation}
for the lower tunnel region, while at the upper one
\begin{equation}\label{eq:QPC2}
\left (\begin{matrix}
\psi_{\uparrow}(t,z=0^-) \\ \psi_{\downarrow}(t,y=\pi R^-) \\ \psi_{\uparrow}(t,y=\pi R^+) \\ \psi_{\downarrow}(t,z=0^+)
\end{matrix}\right )=S_1 \left (\begin{matrix}
\psi_{\downarrow}(t,z=0^-) \\ \psi_{\uparrow}(t,y=\pi R^-) \\ \psi_{\downarrow}(t,y=\pi R^+) \\ \psi_{\uparrow}(t,z=0^+)
\end{matrix}\right )
.\end{equation}
The form of the scattering matrices $S_i$ is~\cite{Inhofer2013proposal, Ferraro2014electronic, Dolcini2011full, Delplace2012magnetic, Dolcetto2012tunneling, Rizzo2013}
\begin{equation}\label{eq:Sqpc}
S_i=\left (\begin{matrix}
0 & p_i & f_i & r_i \\
p_i & 0 & r_i & f_i \\
f_i^* & r_i & 0 & p_i \\
r_i & f_i^* & p_i & 0
\end{matrix}\right )
,\end{equation}
with $p_i$ and $f_i$ the spin-preserving and spin-flipping tunneling amplitudes respectively and $r_i$ the amplitude probability for electrons to remain on the same channel without tunneling at $i$-th QPC. Note that backscattering is forbidden due to TRS, which also implies~\cite{Ferraro2014electronic} $\mathrm{Im}\{r_i\}=\mathrm{Re}\{p_i\}=\mathrm{Re}\{f_i\}=0$.
We denote by $T_i=\vert p_i\vert^2+\vert f_i\vert^2=1-\vert r_i\vert^2$ the total tunneling probability through the $i$-th QPC, with $T=(T_1+T_2)/2$.
By inserting the expressions evaluated from Eq.~\eqref{eq:wavefunctions} into Eq.~\eqref{eq:QPC1} and \eqref{eq:QPC2} allows one to find the scattering amplitudes as a function of the incoming amplitudes $A_{i\sigma}$.
Although it is possible to solve the problem for a more general driving potential~\cite{Moskalets2013}, for simplicity and sake of clarity we only present the solution in the adiabatic regime in which the driving is very slow, which corresponds to calculating the frozen scattering matrix~\cite{Moskalets2011non} with $\Upsilon(t)\approx 1$.
By modelling $U(t)=U_0+U_1\cos(\Omega t+\varphi)$, the adiabatic regime occurs for $2\pi/\Omega\gg \tau$, with $2\pi/\Omega$ the period of the potential ($\varphi$ being a phase shift) and $\tau\approx h/(\Delta T)$ the dwell time spent by the Kramers pair on the antidot.
The constant part of the potential $U_0$ accounts for a detuning of the nearest level in the antidot from the Fermi level, and we assume that the amplitude of the oscillating potential $U_0<U_1<\Delta-U_0$, such that only one antidot level crosses the Fermi level at the resonance times $t_{\pm}=\pm \tfrac{1}{\Omega}\arccos(-\tfrac{U_0}{U_1})-\tfrac{\varphi}{\Omega}$.
This is a necessary condition to inject a single electron and a single hole Kramers pair per cycle, otherwise multiple pairs can be injected.

In the adiabatic regime the outgoing amplitudes are related to the incoming ones as $B_{i\sigma}=\sum_{j\sigma^\prime}\mathcal{S}_{i\sigma}^{j\sigma^\prime}A_{j\sigma^\prime}$, where the frozen scattering matrix $\mathcal{S}$, written in the basis $\{2\uparrow,2\downarrow,1\uparrow,1\downarrow\}$, reads
\begin{widetext}
\begin{equation}\label{eq:Smatrix}
\mathcal{S}=\frac{1}{1-e^{2\pi ikR}r_1r_2}\left (\begin{matrix}
r_2-e^{2\pi ikR}r_1 & 0 & e^{\pi ikR}\left (p_1p_2+f_1f_2\right ) & e^{\pi ikR}\left (p_1f_2-f_1p_2\right )\\
0 & r_2-e^{2\pi ikR}r_1 & -e^{\pi ikR}\left (p_1f_2-f_1p_2\right ) & e^{\pi ikR}\left (p_1p_2+f_1f_2\right )\\
e^{\pi ikR}\left (p_1p_2+f_1f_2\right ) & -e^{\pi ikR}\left (p_1f_2-f_1p_2\right ) & r_1-e^{2\pi ikR}r_2 & 0\\
e^{\pi ikR}\left (p_1f_2-f_1p_2\right ) & e^{\pi ikR}\left (p_1p_2+f_1f_2\right ) & 0 & r_1-e^{2\pi ikR}r_2
\end{matrix}\right )
.\end{equation}
\end{widetext}
A few comments are in order concerning Eq.~\eqref{eq:Smatrix}.
The zeros correspond to the absence of backscattering, which is guaranteed as long as TRS is preserved. Moreover it is easy to check that $B_{1\sigma}\leftrightarrow B_{2\sigma}$ under the exchange $1\leftrightarrow 2$ in the tunneling parameters. In the limit $T_1=0$ and $T_2\neq 0$ (or vice versa) the antidot is coupled to a single edge~\cite{Hofer2013emission, Inhofer2013proposal, Ferraro2014electronic} and all the off-diagonal matrix elements vanish.
Finally, if only one type of tunneling processes is possible (either $p_i=0$ or $f_i=0$), the matrix elements connecting incoming and outgoing states with different spin vanish as expected.

The frozen scattering matrix still implicitly depends on time through the phase factors containing $kR$, after making the replacement~\cite{Moskalets2011non} $k\to k-eU(t)/v$.

For small transmission probability $T$ the antidot energy levels are well resolved and the scattering matrix $\mathcal{S}$ deviates from unity only around resonance~\cite{Moskalets2011non}.
We can therefore expand the matrix elements of $\mathcal{S}$ around the resonance times $t_{\pm}$ to lowest order in the tunneling amplitudes as
\begin{widetext}
\begin{equation}\label{eq:Smatrix_ad}
\mathcal{S}\approx\sum_{\alpha=\pm}\frac{1}{t-t_{\alpha}+i\alpha\gamma_+}\left (\begin{matrix}
t-t_\alpha+i\alpha\gamma_- & 0 & i\alpha\gamma_+\frac{p_1p_2+f_1f_2}{T} & i\alpha\gamma_+\frac{p_1f_2-f_1p_2}{T}\\
0 & t-t_\alpha+i\alpha\gamma_- & -i\alpha\gamma_+\frac{p_1f_2-f_1p_2}{T} & i\alpha\gamma_+\frac{p_1p_2+f_1f_2}{T}\\
i\alpha\gamma_+\frac{p_1p_2+f_1f_2}{T} & -i\alpha\gamma_+\frac{p_1f_2-f_1p_2}{T} & t-t_\alpha-i\alpha\gamma_- & 0\\
i\alpha\gamma_+\frac{p_1f_2-f_1p_2}{T} & i\alpha\gamma_+\frac{p_1p_2+f_1f_2}{T} & 0 & t-t_\alpha-i\alpha\gamma_-
\end{matrix}\right )
,\end{equation}
\end{widetext}
with $\gamma_{\pm}=\gamma_1\pm\gamma_2$, and $\gamma_{1,2}=T_{1,2}/(2M\Omega)$ corresponding to the inverse tunneling rate through the $i$-th barrier, with $M=2\pi \vert e\vert\Delta^{-1}\sqrt{U_1^2-U_0^2}$.

\subsection{Current}
In order to characterize the device we compute the emitted current.
The current injected into the spin $\sigma$ channel through the $i$-th tunnel region is obtained from Eq.~\eqref{eq:Smatrix_ad} as~\cite{Moskalets2011non}
\begin{equation}\label{eq:Iisigma}
I_{i\sigma}(t)=\frac{ie}{2\pi}\sum_{j\sigma^\prime}\int dE \left (\partial_Ef_0\right )\mathcal{S}_{i\sigma,j\sigma^\prime}(t)\partial_t\mathcal{S}_{i\sigma,j\sigma^\prime}^*(t)
,\end{equation}
where $f_0(E)$ is the Fermi distribution function.
At low temperature, the derivative of the Fermi distribution implies that Eq.~\eqref{eq:Iisigma} is evaluated at the Fermi energy.
After straightforward algebra one finds
\begin{equation}\label{eq:Ii}
I_{i\sigma}(t)=-\frac{2\gamma_i}{\pi}\sum_{\alpha=\pm}\frac{\alpha e}{\left (t-t_{\alpha}\right )^2+\gamma_+^2}
,\end{equation}
where TRS implies $I_{i\uparrow}=I_{i\downarrow}$.
By integrating Eq.~\eqref{eq:Ii} over time and summing over the spin degree of freedom one finds that a charge $Q_i=2e\gamma_i/\gamma_+$ is injected through the $i$-th barrier during a time interval $\approx\gamma_+$ around $t_-$, while the same charge is adsorbed around $t_+$.
Therefore the total charge injected by the MC around $t_-$, when the antidot level is driven above the Fermi energy, is $Q=Q_1+Q_2=2e$, corresponding to the emission of exactly one Kramers pair. On the other hand, when the antidot level is pushed below the Fermi energy around $t_+$ an opposite charge $-2e$ is injected, the antidot adsorbing a pair of electrons from the edges.

\subsection{Noise}
To characterize the device we now study on the current-current correlations between the different channels.
In particular we are interested in the zero-frequency symmetrized noise spectral power~\cite{Blanter2000} ($\delta I_{i\sigma}=I_{i\sigma}-\langle I_{i\sigma}\rangle$)
\begin{eqnarray}\label{eq:noise0}
\mathcal{P}_{i\sigma,j\sigma^\prime}&=&\frac{1}{2}\int_0^{2\pi/\Omega}\frac{dt}{2\pi/\Omega}\int_{-\infty}^{\infty}d\tau\left\langle\delta I_{i\sigma}(t)\delta I_{j\sigma^\prime}(t+\tau)\right .\nonumber\\
&+&\left.\delta I_{j\sigma^\prime}(t+\tau)\delta I_{i\sigma}(t)\right\rangle
.\end{eqnarray}
By neglecting the thermal contribution, which is valid if $k_BT\ll \Omega$, only the shot noise contributes, and in the adiabatic regime Eq.~\eqref{eq:noise0} can be evaluated from Eq.~\eqref{eq:Smatrix_ad} as~\cite{Moskalets2011non, Ol2008}
\begin{eqnarray}\label{eq:noise_def}
\mathcal{P}_{i\sigma,j\sigma^\prime}&=&\frac{e^2\Omega}{4\pi}\sum_{q=-\infty}^{\infty}\vert q\vert\sum_{\eta\sigma_1}\sum_{\delta\sigma_2}\left \{\mathcal{S}_{i\sigma,\eta\sigma_1}\mathcal{S}^*_{i\sigma,\delta\sigma_2}\right \}_q\nonumber\\
&\times &\left \{\mathcal{S}^*_{j\sigma^\prime,\eta\sigma_1}\mathcal{S}_{j\sigma^\prime,\delta\sigma_2}\right \}_{-q}
,\end{eqnarray}
where curly braces denote the Fourier transform, $\{\dots\}_q=\Omega/(2\pi)\int_0^{2\pi/\Omega}dte^{iq\Omega t}\{\dots\}$.
Electron and hole emissions contribute independently, $\mathcal{P}_{i\sigma,j\sigma^\prime}=\mathcal{P}_{i\sigma,j\sigma^\prime}^{(e)}+\mathcal{P}_{i\sigma,j\sigma^\prime}^{(h)}$, and are equal, $\mathcal{P}_{i\sigma,j\sigma^\prime}^{(e)}=\mathcal{P}_{i\sigma,j\sigma^\prime}^{(h)}$. After lengthy but straightforward algebra we obtain~\cite{note_ad}
\begin{widetext}
\begin{equation}\label{eq:noise}
\mathcal{P}=\frac{e^2\Omega}{4\pi}\left (\begin{matrix}
\frac{T_1T_2}{T^2} & 0 & -\left (\frac{p_1p_2+f_1f_2}{T}\right )^2 & -\left (\frac{p_1f_2-f_1p_2}{T}\right )^2\\
0 & \frac{T_1T_2}{T^2} & -\left (\frac{p_1f_2-f_1p_2}{T}\right )^2 & -\left (\frac{p_1p_2+f_1f_2}{T}\right )^2\\
-\left (\frac{p_1p_2+f_1f_2}{T}\right )^2 & -\left (\frac{p_1f_2-f_1p_2}{T}\right )^2 & \frac{T_1T_2}{T^2} & 0\\
-\left (\frac{p_1f_2-f_1p_2}{T}\right )^2 & -\left (\frac{p_1p_2+f_1f_2}{T}\right )^2 & 0 & \frac{T_1T_2}{T^2}
\end{matrix}\right )
.\end{equation}
\end{widetext}
Firstly, we notice that a sum rule holds for the zero-frequency noise, $\sum_{j\sigma\prime}\mathcal{P}_{i\sigma,j\sigma^\prime}=0$.
Moreover $\mathcal{P}_{i\sigma,i\bar{\sigma}}=0$ as a consequence of TRS preventing scattering within the Kramers pairs on the same edge: the latter leads to $\langle I_{i\sigma}I_{i\bar{\sigma}}\rangle=\langle I_{i\sigma}\rangle\langle I_{i\bar{\sigma}}\rangle$, which by virtue of Eq.~\eqref{eq:noise0} corresponds to absence of correlation.
The other correlation functions are in general different from zero.
However in the presence of a single type of scattering (either $p_i=0$ or $f_i=0$), channels with opposite spin are uncorrelated, so that $\mathcal{P}_{i\sigma,\bar{i}\bar{\sigma}}=0$.
Finally we note that all the matrix elements vanish if the antidot is coupled to one edge only ($T_i=0$). In this case the Kramers pair can be emitted only in one edge with opposite spin, due to TRS. Therefore, there is no uncertainty in the emitted state and the electron source is noiseless.

\section{Entanglement}\label{sec:entanglement}
Now that we have characterized the MC, showing that it is able to emit exactly one Kramers pair per cycle, we can investigate weather it is able to produce entangled states. To generate entangled two-particle states via Kramers pair injection in 2D TIs two main strategies have been adopted: either locally breaking TRS by piercing the MC with a magnetic field and studying a new type of time-bin entanglement~\cite{Hofer2013emission}, or adding additional QPCs, such that new quantum states can be created and manipulated after the unentangled two-particle state is injected from the MC~\cite{Inhofer2013proposal, Strom2015controllable}.

Contrary to these proposals, the multiple tunneling processes from the MC to the different edges which exist in the setup of Fig.~\ref{fig:antidotQSHdouble} lead to a richer scenario for the investigation of entanglement properties.
Indeed, the emitted two-particle state is in general a superposition of six orthogonal quantum states, depending on the nature of the injection process
\begin{eqnarray}\label{eq:psiTL}
\vert \psi\rangle&=&\frac{1}{2T}\left [T_1c^{\dagger}_{1\downarrow}c^{\dagger}_{1\uparrow}+T_2c^{\dagger}_{2\downarrow}c^{\dagger}_{2\uparrow}\right .\nonumber\\
&+&\left (p_1f_2-f_1p_2\right )\left (c^{\dagger}_{1\uparrow}c^{\dagger}_{2\uparrow}+c^{\dagger}_{1\downarrow}c^{\dagger}_{2\downarrow}\right )\nonumber\\
&+&\left .\left (p_1p_2+f_1f_2\right )\left (c^{\dagger}_{1\uparrow}c^{\dagger}_{2\downarrow}-c^{\dagger}_{1\downarrow}c^{\dagger}_{2\uparrow}\right )\right ]\vert 0\rangle
.\end{eqnarray}
Here, the operator $c^{\dagger}_{i\sigma}$ creates an electron in the $i$-th edge with spin $\sigma$.
The first two states correspond to the Kramers pair being injected into the same edge. The third and fourth states are achieved when the Kramers pair is split into different edges and the two injected particles have the same spin. Finally, the last two states are created when the two particles are split into different edges and have opposite spin.
Note that because of the Pauli principle, Eq.~\eqref{eq:psiTL} does not contain states where both electrons of a Kramers pair are injected into the same edge with the same spin.

Each of the six realizations of $\vert\psi\rangle$ gives rise to a distinct current signal measured by the two contacts in Fig.~\ref{fig:antidotQSHdouble}.
Since TRS prevents backscattering on each edge, the scattering matrix connecting the creation operators at the QPCs $c^{\dagger}_{i\sigma}$ (with $i=1,2$) to the creation operators in the contacts $c^{\dagger}_{\alpha\sigma}$ (with $\alpha=$ L,R) is simply given by
\begin{equation}\label{eq:Sdet}
\left (\begin{matrix}
c^{\dagger}_{\mathrm{L\downarrow}} \\ c^{\dagger}_{\mathrm{R\uparrow}} \\ c^{\dagger}_{\mathrm{R\downarrow}} \\ c^{\dagger}_{\mathrm{L\uparrow}}\end{matrix}\right )=\left (\begin{matrix}
e^{ik_{\mathrm{F}}l_1} & 0 & 0 & 0 \\
0 & e^{ik_{\mathrm{F}}l_2} & 0 & 0 \\
0 & 0 & e^{ik_{\mathrm{F}}l_3} & 0 \\
0 & 0 & 0 & e^{ik_{\mathrm{F}}l_4}
\end{matrix}\right )
\left (\begin{matrix}
c^{\dagger}_{\mathrm{1\downarrow}} \\ c^{\dagger}_{\mathrm{1\uparrow}} \\ c^{\dagger}_{\mathrm{2\downarrow}} \\ c^{\dagger}_{\mathrm{2\uparrow}}\end{matrix}\right )
,\end{equation}
where $l_i$ are the lengths of the different arms of the setup and $k_{\mathrm{F}}$ the Fermi momentum.
Therefore, we can rewrite the final state as
\begin{align}\label{eq:psiTLcont}
    \vert \psi\rangle
&=
    \frac{1}{2T}\bigg[
    T_1 e^{-ik_{\mathrm{F}}(l_1+l_2)} c^{\dagger}_{\mathrm{L}\downarrow}c^{\dagger}_{\mathrm{R}\uparrow}
+
    T_2 e^{-ik_{\mathrm{F}}(l_3+l_4)} c^{\dagger}_{\mathrm{R}\downarrow}c^{\dagger}_{\mathrm{L}\uparrow} \nonumber \\
&+
    \left (p_1f_2-f_1p_2\right )
    \Big(
    e^{-ik_{\mathrm{F}}(l_2+l_4)} c^{\dagger}_{\mathrm{R}\uparrow} c^{\dagger}_{\mathrm{L}\uparrow} \notag \\
&+
    e^{-ik_{\mathrm{F}}(l_1+l_3)} c^{\dagger}_{\mathrm{L}\downarrow} c^{\dagger}_{\mathrm{R}\downarrow}
    \Big) \nonumber\\
&+
    \left (p_1p_2+f_1f_2\right) \Big(
    e^{-ik_{\mathrm{F}}(l_2+l_3)} c^{\dagger}_{\mathrm{R}\uparrow} c^{\dagger}_{\mathrm{R}\downarrow} \notag \\
&-
    e^{-ik_{\mathrm{F}}(l_1+l_4)} c^{\dagger}_{\mathrm{L}\downarrow} c^{\dagger}_{\mathrm{L}\uparrow}
    \Big)
    \bigg]\vert 0\rangle
.\end{align}
The first four terms in Eq.~\eqref{eq:psiTL} give rise to the same number of electrons collected at the left and right detectors, exactly one in each of them. Since normal contacts cannot resolve the spin of the incoming electrons, the fact that each detector collects exactly one electron is associated to an entangled state.
On the other hand, the collection of two electrons in the right (left) contact identifies the fifth (sixth) state, so that no entanglement is present in this case.
Therefore, we project the state in Eq.~\eqref{eq:psiTLcont} to the subspace in which each of the two detectors in Fig.~\ref{fig:antidotQSHdouble} collects exactly one electron. Such a process is referred to as postselection,\cite{Bose, Lebedev} and the postselected state can be written in the standard basis for a spin $\frac{1}{2}$ two-qubit system as
\begin{align}\label{eq:psiTLps}
&
    \vert \tilde{\psi}\rangle
=
    \frac{1}{\tilde{2T}} \bigg[
    T_1e^{-ik_{\mathrm{F}}(l_1+l_2)} \vert\downarrow,\uparrow\rangle
-
    T_2 e^{-ik_{\mathrm{F}}(l_3+l_4)} \vert\downarrow,\uparrow\rangle \nonumber\\
&+
    \left (p_1f_2-f_1p_2\right )
    \Big(
    e^{-ik_{\mathrm{F}}(l_1+l_3)} \vert\downarrow,\downarrow\rangle -
    e^{-ik_{\mathrm{F}}(l_2+l_4)} \vert\uparrow,\uparrow\rangle \Big )
,\end{align}
where we have defined $\vert\sigma,\sigma^\prime\rangle\equiv c^{\dagger}_{\mathrm{L}\sigma}c^{\dagger}_{\mathrm{R}\sigma^\prime}\vert 0\rangle$.
The normalization parameter $\tilde{T}$ is chosen such that $\langle\tilde{\psi}\vert\tilde{\psi}\rangle=1$ and therefore reads $\tilde{T}=T\sqrt{1-\eta/2}$, with
\begin{equation}\label{eq:eta}
\eta=\left (\frac{p_1p_2+f_1f_2}{T}\right )^2
.\end{equation}

\subsection{Concurrence}
\begin{figure}[t]
\centering
\includegraphics[width=\columnwidth]{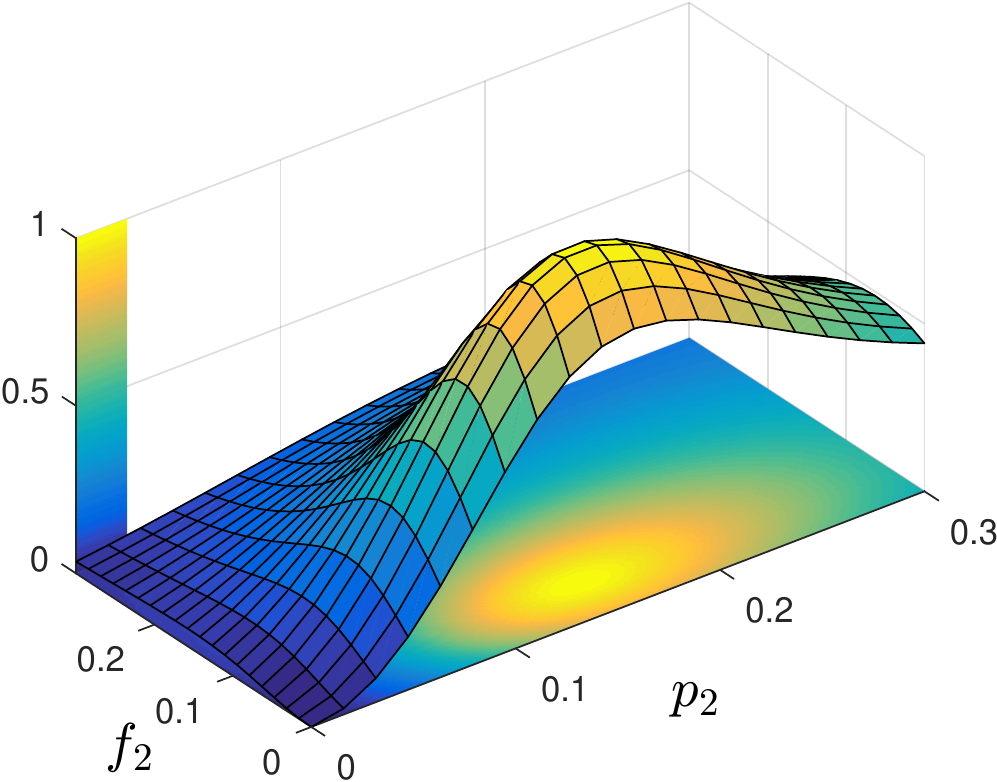}
\caption{(Color online) Plot of the concurrence $\mathcal{C}$ as a function of the parameters of the second tunnel junction $p_2$ and $f_2$, with fixed values $p_1=0.15$ and $f_1=0.05$.}\label{fig:concurrence}
\end{figure}

We use the concurrence $\mathcal{C}$ as a measure of the entanglement~\cite{Wootters}. For a pure state of a bipartite system, it reads
\begin{equation}
\mathcal{C}=\vert\langle\tilde{\psi}\vert\sigma_y\otimes\sigma_y\vert\tilde{\psi}^*\rangle\vert
,\end{equation}
where the complex conjugation is to be taken in the basis~\cite{Strom2015controllable, Wootters} $\{ \vert\sigma,\sigma^\prime\rangle\}$.
Simple algebra gives the expression for the concurrence for the postselected state $\vert\tilde{\psi}\rangle$ in Eq.~\eqref{eq:psiTLps} as
\begin{equation}\label{eq:concurrence}
\mathcal{C}=\frac{\eta}{2-\eta}
.\end{equation}
It is worth pointing out that the concurrence does not depend on the geometrical parameters $l_i$. In this sense, the entanglement is insensitive to the detailed geometry of the device in Fig.~\ref{fig:antidotQSHdouble}, and is not affected, for instance, by the antidot being closer to one of the two contacts.
This is due to the weak influence of dephasing mechanisms on the properties of the helical edge states. Indeed, in the presence of dephasing, additional (random) phases would appear in front of each quantum state in Eq.~\eqref{eq:psiTLps}, which would average out thus producing a final separable (not entangled) quantum state. Experiments in InAs/GaSb reported~\cite{Knez2015robust} phase coherence lengths of around $4.4$ $\mu$m, which should be considered as an upper bound for the size of the device in order to guarantee phase-coherence throughout the system.
The concurrence is invariant under the exchange $1\leftrightarrow 2$.
Quite remarkably, it only depends on the tunneling processes through the combination $\eta$ defined in Eq.~\eqref{eq:eta}.
Note that the fact that $0\leq\eta\leq 1$ implies that $0\leq\mathcal{C}\leq 1$, i.e., the state $\vert \tilde{\psi}\rangle$ can vary between a separable state ($\mathcal{C}=0$) and a maximally entangled one ($\mathcal{C}=1$) depending on the tunneling amplitudes $p_i$ and $f_i$.
A plot of the concurrence $\mathcal{C}$ as a function of the tunneling amplitudes to the lower edge, while keeping fixed those to upper edge, is shown in Fig.~\ref{fig:concurrence}.

Form Eq.~\eqref{eq:concurrence} we see that the state is not entangled if $\eta=0$.
This corresponds to the case where only one edge, say the upper edge, is coupled to the MC: $\mathcal{C}=0$ for $p_2=f_2=0$.
Indeed, in this case we find from Eq.~\eqref{eq:psiTLps} that $\vert\tilde{\psi}\rangle\propto c^{\dagger}_{\mathrm{R}\uparrow}c^{\dagger}_{\mathrm{L}\downarrow}\vert 0\rangle$ is a separable state.
Analogously, we find an unentangled state for $p_{1}=f_{2}=0$ or $f_1=p_2=0$.

On the other hand, a nonzero concurrence $\mathcal{C}> 0$ is generally found when the antidot is coupled to both edges. The maximum value $\mathcal{C}=1$ is reached for $\eta=1$, i.e., if the two tunnel barriers are symmetric ($p_1=p_2$ and $f_1=f_2$), see Fig.~\ref{fig:concurrence}. Indeed, for this symmetric choice of parameters, we find $\vert\tilde{\psi}\rangle\propto \vert\!\downarrow,\uparrow\rangle+e^{i\chi}\vert\!\uparrow,\downarrow\rangle$, which is maximally entangled ($\mathcal{C}=1$) independently of the phase $\chi$.

\subsection{Efficiency}
Figure~\ref{fig:concurrence} shows that with a suitable choice of the tunneling parameters the state $\vert\tilde{\psi}\rangle$ can be entangled, and maximum entanglement $\mathcal{C}=1$ is achieved for symmetric tunneling contacts between the antidot and the edges.
However, as the state $\vert\tilde{\psi}\rangle$ is the result of a postselection procedure, it is important to quantify the efficiency $\mathcal{F}$ of the setup, i.e., to calculate the percentage of states which give rise to exactly one electron at each detector compared to the discarded ones, where both injected electrons are collected at the same detector.
Indeed, in some proposals~\cite{Inhofer2013proposal} maximum entanglement $\mathcal{C}\to 1$ is only possible in the limit of vanishing efficiency $\mathcal{F}\to 0$, meaning that very few injected states can be used to generate entanglement.
Even though different devices have been proposed, the efficiency at maximum entanglement is predicted to be rather small~\cite{Strom2015controllable} $\mathcal{F}\approx 6\%$.
Therefore we will evaluate $\mathcal{F}$ for the setup in Fig.~\ref{fig:antidotQSHdouble} and compare it with the previous proposals. The efficiency can be evaluated from Eq.~\eqref{eq:psiTLcont}, in which the last line corresponds to final states discarded by the postselection process. One finds
\begin{equation}\label{eq:efficiency}
\mathcal{F}=1-\frac{\eta}{2}
.\end{equation}

\begin{figure}[t]
\centering
\includegraphics[width=\columnwidth]{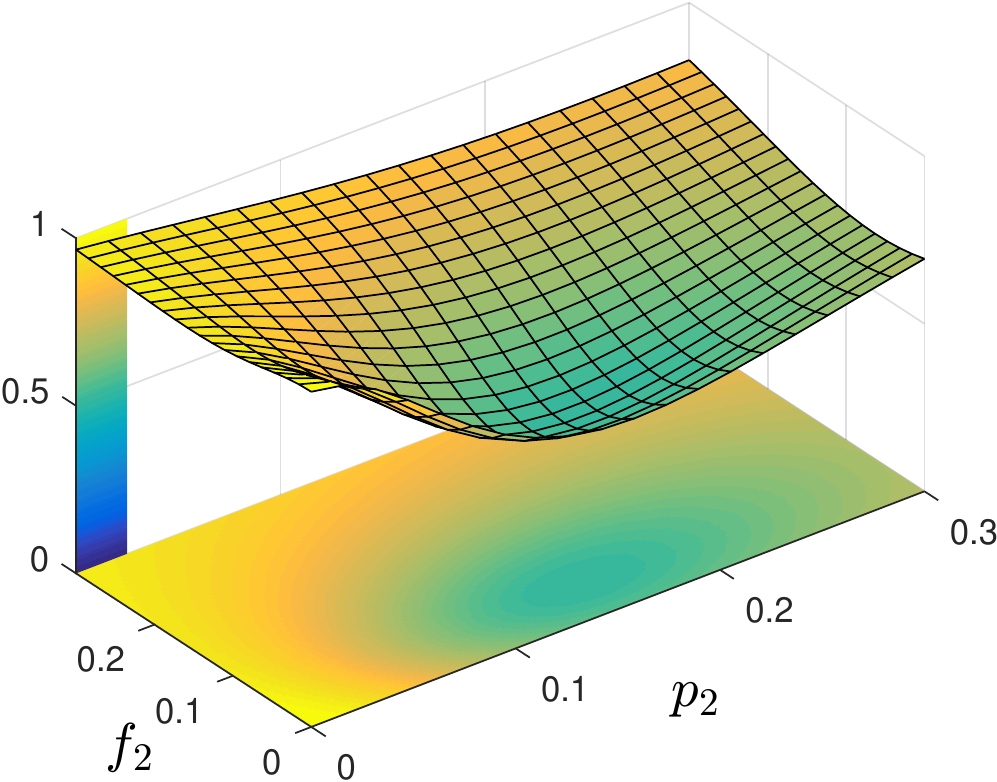}
\caption{(Color online) Plot of the efficiency $\mathcal{F}$ as a function of the parameters of the second tunnel junction $p_2$ and $f_2$, with fixed values $p_1=0.15$ and $f_1=0.05$.}\label{fig:efficiency}
\end{figure}

As the concurrence, the efficiency also depends only on the tunneling parameters via the combination $\eta$ in Eq.~\eqref{eq:eta}.
A plot of $\mathcal{F}$ as a function of the parameters of the second tunnel junction is shown in Fig.~\ref{fig:efficiency}.

Remarkably, the bounds on the parameter $\eta$ imply that $0.5\leq\mathcal{F}\leq 1$.
Maximum efficiency ($\mathcal{F}=1$) is achieved for $\eta=0$. This case, however, corresponds to the absence of entanglement.
This limit is reached, e.g., if $p_2=f_2=0$ or if $p_1=f_2=0$. In these cases, the two electrons are indeed always injected into counter-propagating channels, so that one electron is collected in each detector.

We find that we cannot achieve maximum efficiency $\mathcal{F}=1$ with perfect entanglement $\mathcal{C}=1$.
This can be better seen by rewriting Eqs.~\eqref{eq:concurrence} and $\eqref{eq:efficiency}$ as
\begin{equation}\label{eq:CF}
\mathcal{F}=\frac{1}{1+\mathcal{C}}
,\end{equation}
which shows that maximum efficiency can only be achieved in the absence of entanglement.
However, Eq.~\eqref{eq:CF} also shows that even for maximum entanglement ($\mathcal{C}=1$) the efficiency remains as high as $\mathcal{F}=0.5$, so that a large fraction $\approx 50\%$ of the emitted states is maximally entangled. This indeed corresponds to theoretical maximum limit for non-interacting electrons~\cite{Beenakker2006} and should be compared to previous proposals, where only a fraction $\approx 6\%$ of emitted states was found to be entangled~\cite{Strom2015controllable}.

\subsection{Zero-frequency noise as a measure of entanglement}\label{sec:noise}
The entanglement produced by the device can be measured in the zero-frequency noise~\cite{Samuelsson2003orbital}.
In the two-terminal configuration of Fig.~\ref{fig:antidotQSHdouble} we can define the current-current correlations ($\alpha=$L,R)
\begin{eqnarray}\label{eq:noise_2t}
\mathcal{P}_{\alpha,\beta}&=&\frac{1}{2}\int_0^{2\pi/\Omega}\frac{dt}{2\pi/\Omega}\int_{-\infty}^{\infty}d\tau\left\langle\delta I_{\alpha}(t)\delta I_{\beta}(t+\tau)\right .\nonumber\\
&+&\left.\delta I_{\beta}(t+\tau)\delta I_{\alpha}(t)\right\rangle
,\end{eqnarray}
where for the geometry of Fig.~\ref{fig:antidotQSHdouble} the currents are given by $I_{\mathrm{L}}=I_{1\downarrow}+I_{2{\uparrow}}$ and $I_{\mathrm{R}}=I_{1\uparrow}+I_{2\downarrow}$.
This, together with the multi-channel current-current correlations defined in Eq.~\eqref{eq:noise_def}, allows us to express $\mathcal{P}_{\alpha,\beta}$ in terms of $\mathcal{P}_{i\sigma,j\sigma^\prime}$. For instance, $\mathcal{P}_{R,R}=\mathcal{P}_{1\uparrow,1\uparrow}+\mathcal{P}_{1\uparrow,2\downarrow}+\mathcal{P}_{2\downarrow,1\uparrow}+\mathcal{P}_{2\downarrow,2\downarrow}$.
In particular one finds $\mathcal{P}_{\mathrm{R,R}}=\mathcal{P}_{\mathrm{L,L}}=-\mathcal{P}_{\mathrm{R,L}}=-\mathcal{P}_{\mathrm{L,R}}=\mathcal{P}_0$, with $\mathcal{P}_0=e^2\Omega\eta/2\pi$. By recalling Eqs.~\eqref{eq:concurrence}-\eqref{eq:CF} one finds~\cite{note1}
\begin{equation}\label{eq:noise_FC}
\mathcal{P}_0=\frac{e^2\Omega}{\pi}\mathcal{FC}.
\end{equation}
Equation~\eqref{eq:noise_FC} establishes a direct proportionality between the noise $\mathcal{P}_0$ produced in the two-terminal setup and the product of efficiency and concurrence $\mathcal{FC}$.
It is exact at zero temperature, and it represents a good approximation~\cite{Moskalets2011non} as long as $k_BT\ll\hbar\Omega$. In contrast, at higher temperatures the thermal noise can become the dominant contribution, so that the quantities $\mathcal{P}_{\alpha,\beta}$ can overestimate the entanglement production. In this sense, Eq.~\eqref{eq:noise_FC} should be regarded as the \textit{excess} noise, defined as the difference between the noise measured when the source is on and the noise measured when the source is off, the latter due to thermal fluctuations only. Although the noise is affected by thermal noise, thus corrupting the estimate of entanglement production, the excess noise is not, and represents a more reliable entanglement measurement~\cite{Hofer2013emission}. 

\begin{figure}[t]
\centering
\includegraphics[width=\columnwidth]{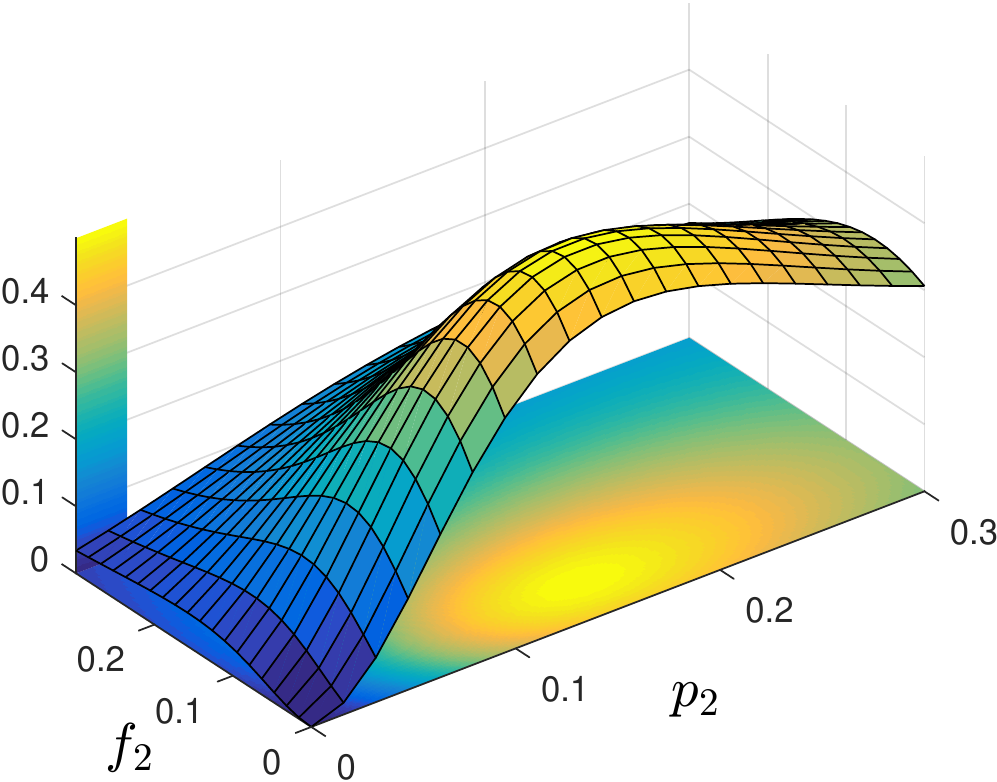}
\caption{(Color online) Plot of the two-terminal noise $\mathcal{P}_0$ in units of $e^2\Omega/\pi$ as a function of the parameters of the second tunnel junction $p_2$ and $f_2$, with fixed values $p_1=0.15$ and $f_1=0.05$.}\label{fig:conceff}
\end{figure}

The (excess) noise $\mathcal{P}_0$ is shown in Fig.~\ref{fig:conceff}, where a maximum value of $\mathcal{P}_0=e^2\Omega/(2\pi)$ is achieved in the symmetric configuration $p_1=p_2$ and $f_1=f_2$, for which $\mathcal{FC}=0.5$ is maximal.
Combined with Eq.~\eqref{eq:CF}, a measurement of the zero-frequency excess noise spectral power Eq.~\eqref{eq:noise_FC} thus makes it possible to extract both the efficiency $\mathcal{F}$ and the concurrence $\mathcal{C}$ separately. Therefore, a shot noise measurement represents a feasible way of measuring the entanglement generated by the device, provided that the temperature is low enough to distinguish it from the thermal noise.

\subsection{Violation of the CHSH inequality}\label{sec:CHSH}
\begin{figure}[t]
\centering
\includegraphics[width=\columnwidth]{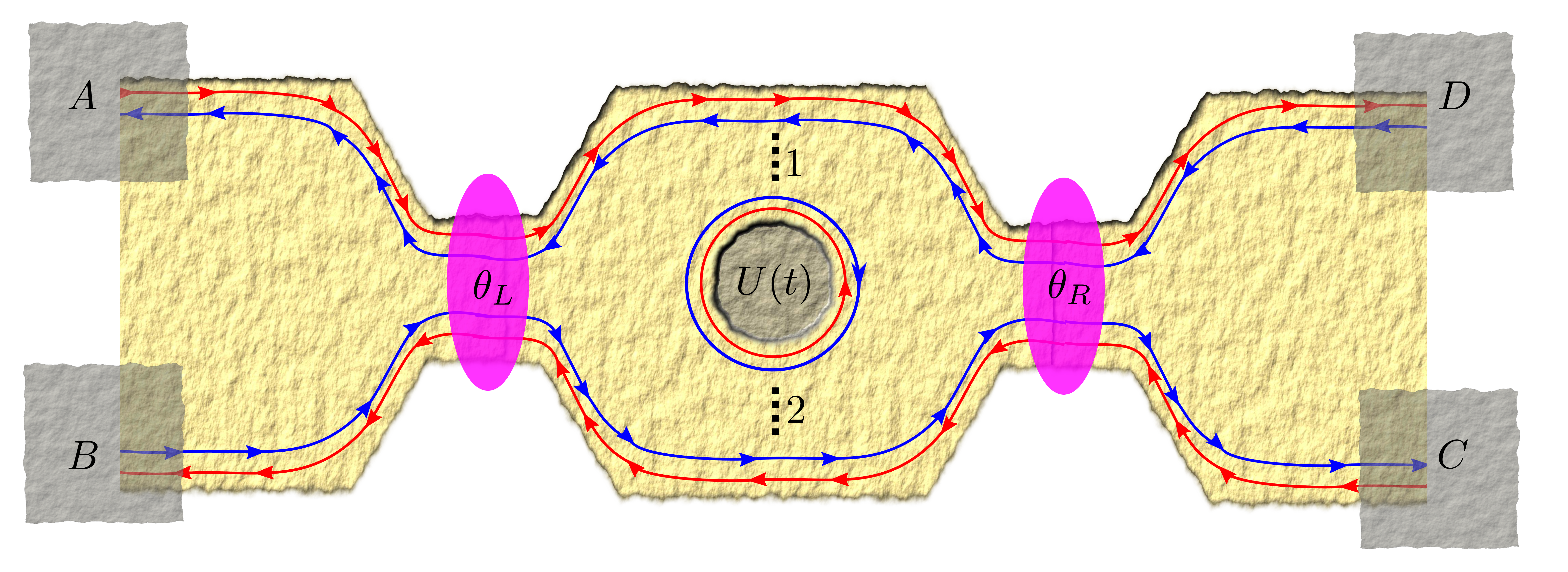}
\caption{(Color online) Setup for detection of the violation of the CHSH inequlity. The injected two-particle state is mixed by two additional QPCs which act as beam splitters; the mixing is tuned via the QPC tunneling parameters $\theta_L$ and $\theta_R$. After the QPCs the particles are collected in a four terminal geometry.}\label{fig:CHSH}
\end{figure}

Alternative schemes to estimate the entanglement production can be considered~\cite{Frustaglia2009, Baltanas2015}.
In particular, it is known that entanglement manifests itself in violations of the Bell inequality, which can therefore be taken as a test for studying the entanglement in the system~\cite{Horodecki2009}.
Specifically, we consider violations of the CHSH inequlity in the setup shown in Fig.~\ref{fig:CHSH}.
The injected Kramers pair propagates towards two additional QPCs, which act as polarizers~\cite{Giovannetti2006, Giovannetti2007}.
In this sense, we focus on the case when only spin-flipping tunneling is possible, which can in principle be realized by properly acting with external gate voltages at the QPCs~\cite{Krueckl2011}; in this case the particles injected from the driven antidot always reach the external contacts, without being backscattered towards the center of the system.
This is parametrized by the scattering matrices
\begin{equation}\label{eq:SLR}
\mathcal{S}_{L/R}=\left (
\begin{matrix}
\cos\theta_{L/R} & \pm i\sin\theta_{L/R} \\
\pm i\sin\theta_{L/R} & \cos\theta_{L/R}
\end{matrix}
\right )
\end{equation}
which allow to connect the states incoming to QPCs to the outgoing ones, collected by the detectors in a four-terminal geometry.
The off-diagonal component in Eq.~\eqref{eq:SLR} represents spin-flipping forward scattering, the diagonal ones representing spin-preserving reflection; therefore $\tan\theta_{L,R}$ represent the ratio between tunneling and reflection amplitude.

The CHSH inequality can be formulated in terms of the normalized particle-number-difference correlators~\cite{Inhofer2013proposal}
\begin{equation}\label{eq:norm-corr}
E(\theta_L,\theta_R)=\frac{\left\langle\left (N_A-N_B\right )\left (N_C-N_D\right )\right\rangle_{\theta_L,\theta_R}}{\left\langle\left (N_A+N_B\right )\left (N_C+N_D\right )\right\rangle_{\theta_L,\theta_R}}
\end{equation}
and reads
\begin{equation}\label{eq:CHSH}
\left\vert E(\theta_L,\theta_R)+E(\theta_L^\prime,\theta_R)+E(\theta_L,\theta_R^\prime)-E(\theta_L^\prime,\theta_R^\prime)\right\vert\leq 2
.\end{equation}
In Eq.~\eqref{eq:norm-corr}, $N_i$ is the particle number operator at terminal $i\in\{A,\dots,D\}$, and the average $\left\langle\dots\right\rangle_{\theta_L,\theta_R}$ is computed on a configuration with the tunneling parameters at the left and right QPCs set to $\theta_L$ and $\theta_R$ respectively.\\
For sake of simplicity, we consider only spin-flipping processes also at the source-edge barriers, that is $p_1=p_2=0$, so that $T_i=\vert f_i\vert^2$.
Explicit evaluation gives
\begin{equation}
E(\theta_L,\theta_R)=\frac{2T_1T_2}{T_1^2+T_2^2}\sin2\theta_L\sin2\theta_R-\cos2\theta_L\cos2\theta_R
.\end{equation}
The CHSH is in general violated, i.e., the left-hand side of Eq.~\eqref{eq:CHSH} is greater than 2, if $T_1T_2\neq 0$, i.e., if the source is coupled to both edges. Indeed in this case the concurrence evaluated from Eqs.~\eqref{eq:eta} and \eqref{eq:concurrence} is $\mathcal{C}>0$, corresponding to an entangled state.
The theoretical maximum value of the left-hand side of Eq.\eqref{eq:CHSH} is $2\sqrt{2}$; this value is achieved for $T_1=T_2$, thus confirming the picture of maximally entangled state $\mathcal{C}=1$ in the case of symmetric source-edge barriers~\cite{thetas}.

We have considered two possible detection schemes for entanglement. In the first case discussed in Sec.~\ref{sec:noise}, entanglement is extracted via zero-frequency noise measurements, whose possible disadvantage is that different sources of noise (such as thermal noise) could lead to an overestimation of the entanglement production.
This problem is overcome by a Bell test of the type considered in Sec.~\ref{sec:CHSH}. However, in this case a fine tuning of the tunneling parameter at the outer QPCs in Fig.~\ref{fig:CHSH} is needed in order to preserve a high efficiency, meaning, the efficiency is reduced by the presence of spin-preserving tunneling at the outer QPCs.
We believe that a combination of these two complementary protocols provides a reliable method to estimate the entanglement production of the single-Kramers pair source.

\section{Conclusions}\label{sec:conclusions}

To summarize, we have investigated the production of entangled electron pairs using an antidot embedded in a two-dimensional topological insulator. The antidot is subject to a time-periodic gate voltage which ensures that it emits or absorbs two electrons per cycle, which can be in an entangled state. In contrast to previous proposals, we have considered a setup where the antidot is coupled by tunneling to two opposite edge of a narrow quantum spin Hall bar. We have found that the emission of electron pairs, together with a postselection procedure, gives rise to entanglement, which can be detected using both measurement of the shot noise in a two-terminal geometry and violation of the CHSH inequality.

We have used the concurrence to quantify the entanglement and investigated the efficiency of the entanglement production process. We have found that our novel proposed setup makes it possible to generate maximally entangled state with an efficiency of $50\%$, significantly higher than the efficiencies achievable in previous proposals where the mesoscopic capacitor was coupled to a single edge channel. Hence, our proposed setup could be used as an efficient source of entangled electrons.

\section*{Acknowledgments}
The authors would like to thank Alexia Rod, Dario Ferraro, and Patrik Recher for helpful discussions. We acknowledge financial support from the DFG priority program SPP 1666 ``Topological insulators'' and from the National Research Fund, Luxembourg under grant ATTRACT 7556175.


\begin{thebibliography}{9}
\bibitem{note_ad} The limit $\sin(\Omega\gamma_+)\approx\Omega\gamma_+$ has been taken Eq.~\eqref{eq:noise}, which is valid in the adiabatic regime.
\end{thebibliography}
\end{document}